\documentclass[english,10ppt,nofootinbib]{revtex4}
\usepackage[T1]{fontenc}
\usepackage[latin9]{inputenc}
\usepackage{amsmath}
\usepackage{amssymb}
\usepackage{graphicx}
\usepackage{color}

\makeatletter
\@ifundefined{textcolor}{}
{%
 \definecolor{BLACK}{gray}{0}
 \definecolor{WHITE}{gray}{1}
 \definecolor{RED}{rgb}{1,0,0}
 \definecolor{GREEN}{rgb}{0,1,0}
 \definecolor{BLUE}{rgb}{0,0,1}
 \definecolor{CYAN}{cmyk}{1,0,0,0}
 \definecolor{MAGENTA}{cmyk}{0,1,0,0}
 \definecolor{YELLOW}{cmyk}{0,0,1,0}
}

\usepackage{babel}

\makeatother

\usepackage{babel}
\begin{document}

\preprint{This line only printed with preprint option}

\title{Analytic Study of Cosmological Perturbations in a Unified Model of Dark Matter and Dark Energy with 
a Sharp Transition}

\author{ Rodrigo R. Cuzinatto$^{1,2}$}
\email{cuzinatto@hep.physics.mcgill.ca}
\author{L\'eo G. Medeiros$^{3,4}$}
\email{leogmedeiros@ect.ufrn.br}
\author{Eduardo M. de Morais$^4$}
\email{emdemor@ift.unesp.br}
\author{Robert H. Brandenberger$^{1}$}
\email{rhb@physics.mcgill.ca}

\affiliation{$^1$Physics Department, McGill University, Montreal, QC, H3A 2T8, Canada}

\affiliation{$^2$Instituto de Ci\^encia e Tecnologia, Universidade Federal de Alfenas,
Rodovia Jos\'e Aur\'elio Vilela, 11999, Cidade Universit\'aria, CEP 37715-400,
Po\c cos de Caldas, MG, Brazil}

\affiliation{$^{3}$Escola de Ci\^encia e Tecnologia, Universidade Federal do Rio Grande
do Norte, Campus Universit\'ario, s/n, CEP 59072-970, Natal, Brazil}

\affiliation{$^4$Instituto de F\'isica Te\'orica, S\~ao Paulo State University, P.O. Box
70532-2, CEP 01156-970 S\~ao Paulo, SP, Brazil}




\begin{abstract}

We study cosmological perturbations in a model of unified dark matter and
dark energy with a sharp transition in the late-time universe. The dark sector
is described by a dark fluid which evolves from an early stage at redshifts
$z > z_C$ when it behaves as cold dark matter (CDM) to a late time 
dark energy (DE) phase ($z < z_C$) when the equation of state
parameter is $w = -1 + \epsilon$, with a constant $\epsilon$ which must be
in the range $0 < \epsilon < 2/3$. We show that fluctuations in the
dark energy phase suffer from an exponential instability, the mode functions
growing both as a function of comoving momentum $k$ and of conformal
time $\eta$. In order that this exponential instability does not lead to
distortions of the energy density power spectrum on scales for which
we have good observational results, the redshift $z_C$ of transition
between the two phases is constrained to be so close to zero that
the model is unable to explain the supernova data. 

\end{abstract}

\maketitle




\section{Introduction \label{sec:Intro}}

Observations of large scale structure \cite{SDSS,WiggleZ} are consistent
with a cosmological model where cold dark matter (CDM) dominates the
evolution of the universe after recombination. CDM produces a decelerating 
expansion of the universe. However, magnitude-redshift studies of 
supernovae of type Ia (SNIa) \cite{Riess1998,Perlmutter1999} indicate that 
the universe's expansion is currently accelerating. This demands the presence 
of an exotic component dubbed dark energy (DE) which has an equation
of state close to that of a cosmological constant and which dominates the
energy density at the present time. This component affects the
detailed peak structure of the angular power spectrum of cosmic 
microwave background radiation (CMB) anisotropies (see e.g.
\cite{Hu2001,Planck2013} and \cite{Mukhanov2005,Dodelson2003} for
textbook discussions). The current data (see e.g. \cite{Planck2015})
give detailed support of the current picture in which the universe is
dominated by CDM until the DE component takes over in the recent past.

The cosmological constant $\Lambda$ is the simplest candidate for
dark energy: it is constant in space and time and has a geometrical justification
in that it is a term which can be added to the Einstein-Hilbert action of General Relativity,
and it fits current observational data available with great accuracy (see e.g. \cite{Li2011}). 
However, positing a cosmological constant as dark energy leads to a number
of conceptual problems (see e.g. \cite{Weinberg1989,Carroll1992}): 
Since the vacuum energy of quantum fields acts as a cosmological constant
and has an expected magnitude which can be estimated to be many orders of
magnitude larger than the currently observed dark energy density, it leads to
the problem of not knowing what causes the cosmological constant to be so small.
This is the ``old'' cosmological constant. If we had an explanation for a suppressed
value, we would still have to explain the coincidence that the cosmological
constant begins to rear its head at the present time. This is the ``coincidence''
problem. Hence, it is of interest to pursue other explanations for the
presence of dark energy \footnote{Note that even if dark energy is a new component
of matter, the ``old'' cosmological constant problem remains.}.

Alternatives to a cosmological constant as dark energy (the so-called
$\Lambda$CDM model) have been explored
as possible explanations for the present-day acceleration while preserving the
conditions for structure formation in the young universe. A class of such
scenarios is based on adding a new component of matter, e.g. quintessence \cite{Ratra1988,Wetterich,Sola,Caldwell1998,Carroll1998,Tsujikawa2013}, k-essence models \cite{Chiba2000,Picon2001,Bertacca2010} -- both field-theory-based frameworks -- 
or a Chaplygin gas \cite{Kamenshchik2001,Bento2002} -- a single fluid with an exotic 
equation of state -- to name but a few. Constraints to some of these models were 
presented in \cite{Felice2013}. There are also attempts to explain dark energy
using modified gravity (see e.g. \cite{Capozziello2011,Nojiri2011,Faraoni2010}). They
prefer to alter the geometrical equations on large distances instead of
assuming the existence of an exotic matter field or fluid. In both classes of
models, the challenge is to ensure that cosmological fluctuations evolve in
a way consistent with the predictions of the $\Lambda$CDM model. We will
here follow the option of modifying the matter sector of the theory.

To explain the observed background cosmology, what one needs is: (1) two dark components which in turn dominate the energy content of the universe,
namely CDM followed by some type of DE; or, (2) a single component that
evolves from CDM to a $\Lambda$-like behavior. This paper deals
with possibility number (2). In addition, we will make the crucial assumption that
this dark component can be described as a perfect fluid which will have
to have a time-dependent equation of state (EoS) starting with
the EoS parameter $w=0$ of CDM and evolving at late times into a fluid
characterized by $w<-1/3$ to account for DE. We call this a ``unified''
dark fluid model (UDM).

Such models have been explored in the literature under the name ``quartessence''
models -- see e.g.  \cite{Cuzinatto2016,Waga2003,Linder2005,Ribamar2012,Pietrobon2008,Bruni2013,Leanizbarrutia2017} 
and references therein. In particular, the authors in Ref. \cite{Cuzinatto2016} analyzed the 
background cosmology of a model with
\begin{equation}
w(z) = \frac{1}{\pi} \arctan (\alpha z + \beta) -\frac{1}{2} \label{eq:w_arctan}
\end{equation}
and its parameters $(\alpha,\beta)$ were constrained using SNIa \cite{Union}, 
baryon acoustic oscillations \cite{BOSS,6dFGRS}, gamma-ray bursts \cite{GRB} 
and primordial nycleosynthesis \cite{PN} data sets. In this model, the parameter 
$\alpha$ controls the transition rate between the matter dominated phase and the 
period of accelerated expansion; and the parameter $\beta$ is associated 
with the present-day value of EoS parameter $w(0)$. Ratio $z_C=\beta/\alpha$ gives 
the redshift of transition between the two regimes. Fits to the various data sets led to $z_C=0.40\pm0.05$. By using the same data, $\Lambda$CDM gives $z_C=0.43\pm0.05$. The conclusion was that the arctan-UDM and 
$\Lambda$CDM models are statistically equivalent at the level of background analysis.

At the level of background data like the ones mentioned in the previous paragraph,
the quartessence models appear to be as viable as the standard
$\Lambda$CDM model. This is true  for the arctan-UDM of Eq.~(\ref{eq:w_arctan}) 
but also for the ``Generalized Chaplygin Gas'' (GCG) \cite{Bento2002,Amendola2010} which
is defined by the EoS parameter
\begin{equation}
w(z) = \frac{-1}{\beta \left( 1+z \right)^{3 \left( 1+\alpha \right)} + 1} \, . \label{eq:w_GCG} 
\end{equation}

However, an acceptable model for the unified dark sector must also agree with
the data on cosmological perturbations, in particular the power spectrum
of the large-scale structure. This leads to a serious problem for models in
which dark energy is described by a fluid since the negative value of the
equation of state parameter leads to an exponential instability for the evolution
of density perturbations. This difficulty is precisely the one faced by GCG model for 
negative values of the parameter $\alpha$ in Eq. (\ref{eq:w_GCG}) unless $| \alpha | < 10^{-5}$
\cite{Amendola2010,Aurich2017} \footnote{Positive values of the parameter $\alpha$ in 
GCG models lead to spurious oscillations in the small-scales regime of the 
matter power spectrum also if $| \alpha | < 10^{-5}$.}. But for such small values of
$\alpha$ the model is indistinguishable from the concordance $\Lambda$CDM 
model whose effective EoS parameter is Eq. (\ref{eq:w_GCG}) with 
$\alpha = 0$ and $\beta = \left( \Omega_{m0} / \Omega_{\Lambda} \right)$, where 
$\Omega_{m0}$ ($\Omega_{\Lambda}$) is the matter (cosmological constant) density 
parameter as measured today. The purpose of this paper is to study the constraints
for our quartessence models which come from demanding consistency with the
observed density power power spectrum.

In Ref. \cite{Aurich2016} Aurich and Lustig worked out the CMB angular power spectrum 
predicted by the arctan-UDM model proposed in \cite{Cuzinatto2016} and they found a 
parameter range that allows for consistency with the Planck 2015 data 
\cite{Planck2015} \footnote{This is the case if one admits a vanishing effective speed of sound $c_{\text{eff}}^2$.}. Here we will focus on density fluctuations which are more strongly
affected by the instabilities in the dark energy phase. 

Both in the case of CGC and in the arctan-UDM model mentioned above,  the linearized
fluctuation equations were studied mostly numerically. 
The advantage of a numerical approach is the high-accuracy of the results. The
advantage of an analytical analysis is the better understanding of the qualitative
physical behavior which results. This is the approach we take in this work.

To simplify the analytic analysis of the perturbed equations in a UDM model we will 
simplify the functional form of the corresponding UDM EoS parameter $w(z)$. 
Specifically, we consider a piecewise-UDM EoS parameter
\begin{equation}
w(z)=\begin{cases}
-1+\epsilon\,, & 0 < z < z_{C}\\
0\,, & z \geqslant z_{C} 
\end{cases}\label{eq:w_piecewiseUDM}
\end{equation}
with a constant $\epsilon$ in the range
\begin{equation}
0 < \epsilon \leqslant 2/3 \, . \label{eq:epsilon_interval} 
\end{equation}
The value $z = z_C$ marks the sharp transition between a matter-like behavior ($w=0$) at 
hight redshift ($z \geqslant z_C$) and a dark-energy-like regime at recent times 
$0 \leqslant z \leqslant z_C$. For $w = -1+\epsilon$ to be compatible with a DE phase, it 
should satisfy $w < -1/3$, which explains the upper limit for the interval in 
(\ref{eq:epsilon_interval}). The value $w=-1$ corresponds to a cosmological constant. Note 
that $w < -1$ is the phantom regime. Since we want to avoid the phantom regime and
also not describe a pure cosmological constant, we should have $\epsilon > 0$ 
as in the lower limit (\ref{eq:epsilon_interval}). 

Notice that Eq. (\ref{eq:w_piecewiseUDM}) indeed describes a 
UDM model because the single dark fluid evolves from a CDM regime to a DE regime, 
even though this transition is instantaneous. Conversely, in the $\Lambda$CDM model, 
dark matter and dark energy co-exist, even though CDM dominates initially and then 
DE takes over as the main (but not the only) component in the form of $\Lambda$. 
However, to be consistent with observations, our piecewise $w(z)$ should reproduce 
the background dynamics of the $\Lambda$CDM model today. In particular, this implies
that the first line of Eq. (\ref{eq:w_piecewiseUDM}) should reproduce the effective EoS 
parameter of that model at the current time:
\begin{equation}
w^{\Lambda\text{CDM}}_{\text{eff},0} = \frac{\left( p_{\text{CDM},0} + p_{\Lambda} \right)}{ \left( \varepsilon_{\text{CDM},0} + \varepsilon_{\Lambda} \right) } = \frac{-1}{ \left( \Omega_{\text{CDM},0} / \Omega_{\Lambda} \right) + 1} \simeq -0.72\, . \label{w_LCDM_today}
\end{equation}
($p$ is the pressure and $\varepsilon$ is the energy density.) Hence, we should have 
$\epsilon \simeq 0.28$. Notice that $\epsilon$ \emph{is not} negligible, so our 
model's dark energy phase is very different from a pure $\Lambda$ phase.  

In spite of its simplicity, the piecewise constant UDM model can be viewed as a 
limit towards which other UDM models tend to. This is particularly true for the 
case of the arctan-UDM model of Eq. (\ref{eq:w_arctan}). For 
this model, the transition redshift $z_C = \beta/ \alpha$ corresponds to $w(z_C) = - 1/2$ 
and the $\alpha$ parameter controls the rate in which the transition occurs. By taking 
$(\alpha,\beta) \rightarrow \infty$ but keeping $z_C$ finite we guarantee that the 
arctan-UDM EoS parameter undergoes a sharp transition just like the one described 
by our piecewise-UDM equation of state.

The parameters $(\alpha,\beta)$ in the generalized Chaplygin gas $w(z)$ -- 
Eq. (\ref{eq:w_GCG}) -- can also be tuned to produce a brisk transition from
CDM-like behavior to a DE-like pattern, although this choice is more evolved than in 
the arctan-UDM case. This is because the transition rate $v_C$ in a GCG model is 
intertwined with the transition redshift $z_C$. For instance, when one increases 
the value of $\alpha > 0$ to increase the transition slope $v_C$ then $z_C$ approaches zero. 

A third class of UDM models which tends to the piecewise-UDM model at the background 
level is the one proposed by Linder and Huterer \cite{Linder2005} characterized 
by \footnote{Ref. \cite{Reis2016} offers a similar parameterization but in 
terms of the deceleration parameter.}:
\begin{equation}
w(z) = w_f + \frac{w_i - w_f}{1 + \left( \frac{1 + z_t}{1 + z} \right)^{1/d}} \, , \label{eq:w_Linder}
\end{equation} 
where label $i$ ($f$) stand for the \emph{initial} (\emph{final}) values of $w$, 
$z_t = z_C$ is the transition redshift and $d$ is the fourth parameter in the model. 
This model is a generalization of the GCG scenario, which can be recovered by 
letting $w_i = 0$, $w_f = -1$, $(1+z_t)=\beta^d$ and $d=1/3(1+\alpha)$. 
Consequently, a sharp CDM-to-DE transition is not trivially obtained in the Linder-Huterer 
parameterization for $w(z)$. It is nonetheless possible to show there is such a 
transition in this four-parameter model \footnote{In particular, for $w_i = 0$, 
$d$ controls the transition rate while both $d$ and $w_f$ determine $z_C$ which is 
defined as the value setting $w(z_C)=-1/2$. Assuming $z_C \sim 0.5$, one shows
that the abrupt transition occurs for large negative values of $w_f $ such that 
$|w_f| \gg 1/d$. The price to pay for this choice is to end up with a pronounced phantom regime.}.

Given these connections with other UDM parametrizations, it is interesting to use the 
model with the piecewise constant EoS parameter of Eq. (\ref{eq:w_piecewiseUDM}) to 
perform the study of cosmological perturbations induced by the unified dark fluid. This
is the topic of this paper. In  Sect. \ref{sec:PerturbEqs} we set up the
basic equations. The details of the analytic analysis leading to the formulas for
density fluctuations for our model are found in Sect. \ref{sec:PerturbUM}. The
consequences for structure formation in the universe are discussed
in Sect. \ref{sec:PS}. Final comments are given in Sect. \ref{sec:Conclusions}. 

Note that we work in natural units in which the speed of light and Planck's
constant are set to 1. We work in the context of an expanding 
Friedmann-Robertson-Walker-Lemaitre space-time with superimposed
linear cosmological perturbations. Since the equations of motion simplify, we use conformal
time $\eta$ instead of the physical time $t$. The comoving spatial coordinates are
denoted by $x$.




\section{Cosmological Perturbations \label{sec:PerturbEqs}}

We work in longitudinal gauge in which the perturbed line element takes the 
form \cite{Brandenberger1992, Mukhanov2005}
\begin{equation}
ds^{2}=a^{2}\left(  \eta\right)  \left[  \left(  1+2\phi\right)  d\eta
^{2}-\left(  1-2\phi\right)  \delta_{ij}dx^{i}dx^{j}\right]  \label{eq:ds2}
\end{equation}
if only linear scalar fluctuations for a perfect fluid are taken into account. The function 
$a=a(\eta)$ is the scale factor, and the potential $\phi(x, \eta)$ is the generalization of 
Newtonian gravitational potential.

The equation governing the evolution of scalar perturbations for a
barotropic fluid $p=p\left(  \varepsilon\right)  $ is given by
\cite{Brandenberger1992,Brandenberger2004}:%
\begin{equation}
\phi^{\prime\prime}+3\left(  1+c_{s}^{2}\right)  \mathcal{H}\phi^{\prime
}-c_{s}^{2}\nabla^{2}\phi+\left[  2\mathcal{H}^{\prime}+\left(  1+3c_{s}%
^{2}\right)  \mathcal{H}^{2}\right]  \phi=0,\label{eq:DiffEqPert}%
\end{equation}
where a prime indicates the derivative with respect to conformal time $\eta$ and $\mathcal{H}$ is Hubble function in conformal time, i.e. $\mathcal{H}=a^{\prime}/a$. The quantity $c_{s}^{2}\equiv p^{\prime}/\varepsilon^{\prime}$ is the square of the (adiabatic) speed of sound.

We consider a generic parametrization of the equation of state 
\begin{equation}
p=w\left(  z\right)  \varepsilon \, , \label{eq:EOS}
\end{equation}
where the cosmological redshift $z$ is given as usual in terms of 
$a$ as $(1+z)=a_{0}/a$, and one usually normalizes the scale factor
to be one at the present time $\eta_{0}$, i.e. $a(\eta_{0}) = a_{0} = 1$. Given the time dependent equation of state of Eq.~(\ref{eq:EOS}),
\begin{equation}
c_s^2 = w(z) + \frac{1}{3} \frac{\left( 1 + z \right)}{\left[ 1 + w(z) \right]} \frac{\partial w}{\partial z} . \nonumber
\end{equation}

For a constant $w$ perfect fluid matter we have $c_s^2 = w$. Since in the dark energy phase $w < 0$, this leads to an instability in the fluctuation equations. This is not the case for a scalar field model of dark energy such as quintessence when (for a canonically normalized scalar field) $c_s^2 = 1$. 

Taking the Fourier transform of Eq.~(\ref{eq:DiffEqPert})
one obtains:
\begin{align}
\phi_{{\bf k}}^{\prime\prime}+3\left[1+w(z)\right]\mathcal{H}(z)\phi_{{\bf k}}^{\prime}+w(z)\,k^{2}\phi_{{\bf k}}\qquad\qquad\qquad\nonumber \\
+\frac{1}{3}\frac{\left(1+z\right)}{\left[1+w\left(z\right)\right]}\frac{\partial w}{\partial z}\left[3\mathcal{H}\phi_{{\bf k}}^{\prime}+k^{2}\phi_{{\bf k}}+3\mathcal{H}^{2}(z)\phi_{{\bf k}}\right] & =0\,.\label{eq:FourierDiffEqphi(eta,z)}
\end{align}
The expansion rate $\mathcal{H}$ (in conformal time) appearing above
can be expressed as
\begin{equation}
\mathcal{H}(z)=\frac{H_{0}\Omega_{\text{ref}}^{1/2}}{\left(  1+z\right)  }%
\exp\left\{  \frac{3}{2}\intop_{z_{\text{ref}}}^{z}\frac{\left[
1+w(z^{\prime})\right]  }{(1+z^{\prime})}dz^{\prime}\right\}  \,,
\label{eq:calH(z)}
\end{equation}
where $H_0$ is the current expansion rate (in terms of physical time) and $\Omega$
is the ratio of the energy density to the critical energy density for a spatially flat
universe, i.e. $\Omega\equiv\tilde{\varepsilon}/\tilde{\varepsilon}_{c}$
with $\tilde{\varepsilon}_{c}=3H_{0}^{2}/(8\pi G)$ (tilde signs on top of quantities 
like $\varepsilon$ indicate they are calculated in the unperturbed space-time). 
$\Omega_{\text{ref}}$ and $z_{\text{ref}}$ are integration constants to
be conveniently chosen. Eq.~(\ref{eq:calH(z)}) stems from Eq.~(\ref{eq:EOS}) 
and from the background energy conservation equation.

For a general time dependence of the equation of state parameter it is
not easy to find solutions of Eq.~(\ref{eq:FourierDiffEqphi(eta,z)}). This
difficulty can be overcome by adopting a model with an instantaneous transition between
the matter regime ($w=0$) and the dark energy phase, i.e. by taking an EoS modeled by
Eq.~(\ref{eq:w_piecewiseUDM}). We emphasize that $\epsilon=\text{constant}$ so
that even during the DE-phase $\partial w/\partial z$ vanishes along with
the second line of the differential Eq.~(\ref{eq:FourierDiffEqphi(eta,z)}).
This drastically simplifies the analysis, as will be explored in
Sect. \ref{sec:PerturbUM}.

Once we have determined the relativistic potential $\phi_{\mathbf{k}}$ by
solving the differential equation (\ref{eq:FourierDiffEqphi(eta,z)}), the result can be
used to determine the energy density perturbation 
$\delta\varepsilon_{\mathbf{k}}$. From the $00$-component of the perturbed 
Einstein equations in momentum space \cite{Brandenberger1992, Mukhanov2005}
we have
\begin{equation}
\delta_{\mathbf{k}}=\frac{\delta\varepsilon_{\mathbf{k}}}{\tilde{\varepsilon}%
}=-2\left[  \frac{1}{3}\left(  k\mathcal{H}^{-1}\right)  ^{2}+1\right]
\phi_{\mathbf{k}}-\frac{2}{\mathcal{H}}\phi_{\mathbf{k}}^{\prime}\,,
\label{eq:densitycontrast(phi)}%
\end{equation}
where $\delta_{\mathbf{k}}$ is the fractional density contrast.

The power spectrum $P_{\mathbf{{k}}}$ is calculated with $\delta_{\mathbf{k}}$
and then confronted with observations. In order to obtain 
$\delta_{\mathbf{k}}$ -- and $\phi_{\mathbf{k}}$ before it -- we first need to determine 
the background evolution, i.e. the functions $\mathcal{H}\left(  \eta\right)$
and $a\left(  \eta\right)  $. For our piecewise-UDM model this is done with
Eq.~(\ref{eq:calH(z)}) and the constraint $w=\text{const}$. It yields
\begin{equation}
\mathcal{H}^{2} = H_{0}^{2} \frac{ \Omega_{0} }{ (1+z_{\text{ref}})^{2} }
\left(  \frac{1+z}{1+z_{\text{ref}}} \right)  ^{\left(  1+3w\right)  }
\qquad(w=\text{const})\,. \label{eq:calHwconstant}
\end{equation}
The scale factor $a\left(  \eta\right)  $ is then obtained by using the
definition of $\mathcal{H}$ and $z$ in terms of $a$
\begin{equation}
a(\eta) = a_{\text{ref}} \left[  1 + \frac{\left(  1+3w\right)  }{2} H_{0}
a_{\text{ref}} \Omega_{\text{ref}}^{1/2} \left(  \eta- \eta_{\text{ref}}
\right)  \right]  ^{\frac{2}{\left(  1+3w\right)  }}\qquad(w=\text{const})\,.
\label{eq:awconstant}
\end{equation}
Substituting this result into (\ref{eq:calHwconstant}), we have, 
\emph{up to integration constants}:
\begin{equation}
\mathcal{H}(\eta)=\frac{2}{\left(  1+3w\right)  }\frac{1}{\eta}\qquad
(w=\text{const})\,, \label{eq:calH(eta)wconstant}
\end{equation}
so that Eq. (\ref{eq:FourierDiffEqphi(eta,z)}) for $\phi_{\mathbf{{k}}}$ leads
to:
\begin{equation}
\phi_{\mathbf{k}}^{\prime\prime}+\frac{6\left(  1+w\right)  }{\left(
1+3w\right)  }\frac{1}{\eta}\phi_{\mathbf{k}}^{\prime}+w\,k^{2}\phi
_{\mathbf{k}}=0\qquad(w=\text{const})\,. \label{eq:DiffEqphi(eta)wconstant}
\end{equation}

In the following we will study the solutions of this equation. The perturbations in 
our model first evolve like in a dust-dominated universe, and
at some critical time $\eta_{C}$ transit to unstable solutions in the dark energy
phase. Note that subtleties concerning the integration constants missing in 
Eqs.~(\ref{eq:calH(eta)wconstant}) and (\ref{eq:DiffEqphi(eta)wconstant}) are 
discussed in the Appendix \ref{app:Background}, where the background 
cosmology results for our piecewise-UDM model are presented in detail.




\section{Perturbations in the Unified Dark Fluid Model\label{sec:PerturbUM}}

We will study the fluctuations in the two regimes of our model specified in 
Eq.~{\ref{eq:w_piecewiseUDM}}. The fluctuations generated in some primordial
phase first evolve in the CDM phase. There is then an abrupt transition to
the dark energy phase. We will discuss these phases in turn, as well as
the matching of the fluctuation modes at the transition between the phases.


\subsection{Perturbations in the CDM Phase  \label{sec:CDM-Perturb}}

Dust-like matter is characterized by $w=0$ in which case Eq.~(\ref{eq:DiffEqphi(eta)wconstant})
reduces to:
\begin{equation}
\phi_{{\bf k}}^{\prime\prime}+\frac{6}{\eta}\phi_{{\bf k}}^{\prime}=0 \,,\label{eq:DiffEqphi(eta)dust}
\end{equation}
whose solutions are
\begin{equation}
\phi_{{\bf k}} = \bar{C}_{{\bf{k}},1}+\frac{\bar{C}_{{\bf{k}},2}}{\eta^{5}} \, , \label{eq:phikdust}
\end{equation}
where $\bar{C}_{{\bf{k}},1}$ and $\bar{C}_{{\bf{k}},2}$ are integrations constants determined 
from early-universe physics (inflation or alternative scenarios). We will be normalizing them by fitting to the matter power spectrum at some pivot scale -- 
see Sect. \ref{sec:PS}. Eq.~(\ref{eq:phikdust}) exhibits a constant mode and a decaying mode 
(scaling with $\eta^{-5}$) for the gravitational potential. The density contrast $\delta_{\bf{k}}$ 
is calculated by using (\ref{eq:phikdust}) and inserting $\cal{H}$ (Eq.~(\ref{eq:calH(eta)wconstant})) 
into (\ref{eq:densitycontrast(phi)}), resulting in
\begin{equation}
\delta_{\bf{k}} = -2 \bar{C}_{{\bf{k}},1} \left[ \frac{1}{2} \frac{(k \eta)^2}{6} + 1\right] - 3 \bar{C}_{{\bf{k}},2} \left[ \frac{1}{3} \frac{(k \eta)^2}{6} -1 \right] \frac{1}{\eta^5} \, . \label{eq:deltakdust}
\end{equation}

Fluctuations in the energy density are coordinate-dependent, but the gauge
dependence is not important on sub-Hubble scales, i.e. on scales where the
physical wavelength of the fluctuation mode is smaller than the Hubble radius 
$d_H = {\cal{H}}^{-1} $. Scales of observational interest today are initially
super-Hubble. They cross the Hubble radius when $\lambda = d_H$, i.e. 
$k = {\cal{H}}$. Since in the sub-Hubble regime $k\eta \gg 1$, the
matter density contrast (\ref{eq:deltakdust}) in this regime regime can be
approximated by
\begin{equation}
\delta_{\bf{k}} = - \frac{(k \eta)^2}{6} \left( \bar{C}_{{\bf{k}},1} + \frac{ \bar{C}_{{\bf{k}},2} }{ \eta^5 } \right) \qquad (k \eta \gg 1) \, . \label{eq:deltakdustsub}
\end{equation}
This is the equation we will use (together with (\ref{eq:phikdust})) when matching the 
fluctuations to those in the DE phase.


\subsection{Perturbations for the DE Phase \label{sec:DE-Perturb}}

Inserting the DE sector EoS parameter $w = -1 + \epsilon$ into the differential equation (\ref{eq:DiffEqphi(eta)wconstant}) for $\phi_{\bf{k}}$ we obtain
\begin{equation}
\phi_{{\bf k}}^{\prime\prime}+\frac{6\epsilon}{\left(-2+3\epsilon\right)}\frac{1}{\eta}\phi_{{\bf k}}^{\prime}+\left(-1+\epsilon\right)\,k^{2}\phi_{{\bf k}}=0 \,.\label{eq:DiffEqphi(eta)quasiLambda}
\end{equation}
We see that there is an instability for $\epsilon < 1$. This instability affects shorter wavelengths
more than longer wavelength ones. This leads, in addition to an amplification of the
overall amplitude of the spectrum (which could be compensated by a decrease in the
amplitude of the primordial spectrum) to a change in the spectral shape. It is this change
in shape which will lead to the strong constraints on the model.

Note that in the case of DE described by a cosmological constant ($\epsilon = 0$)
it would appear at first sight that there also is an instability leading to 
$\phi_{{\bf k}}=D_{1}e^{k \eta}+D_{2}e^{-k \eta}$. However, the 0$i$-component of 
the Einstein equations for perturbations demands $D_{1} = D_{2}=0$, meaning that 
$\Lambda$ produces no $\phi_{\bf{k}}$, therefore no $\delta_{\bf{k}}$, i.e. 
$\Lambda$ does not agglomerate. In the case of scalar field dark energy, even
though $w < 0$ the speed of sound is positive, and hence there is no instability.

The general solution of Eq.~(\ref{eq:DiffEqphi(eta)quasiLambda}) can be
written in terms of Bessel functions \cite{Gradshteyn2007}:
\begin{equation}
\phi_{\mathbf{k}}\left(\sqrt{w}k\eta\right)=\left(\sqrt{w}k\eta\right)^{-\nu}
\left[C_{{\bf{k}},1}J_{\nu}\left(\sqrt{w}k\eta\right)+C_{{\bf{k}},2}Y_{\nu}\left(\sqrt{w}k\eta\right)\right]\,, \label{eq:phikBessel}
\end{equation}
where
\begin{equation}
\nu\equiv\frac{1}{2}\left(\frac{5+3w}{1+3w}\right)=-\frac{1}{2}\left(\frac{2+3\epsilon}{2-3\epsilon}\right)<0 \, . \label{eq:nuquasiL}
\end{equation}

The solution for $\phi_{\bf{k}}$ in terms of Bessel functions can be simplified to elementary 
functions if we consider perturbation which are sub-Hubble in the DE regime, i.e. 
$k \eta \gg 1$.  Then, the argument $\zeta=\sqrt{w}k\eta$ of the Bessel functions in 
$\phi_{\bf{k}}$ is large and we can use the asymptotic forms \cite{Gradshteyn2007}:
\begin{align*}
J_{\pm\left\vert \nu\right\vert }\left(\zeta\right) & \simeq\sqrt{\frac{2}{\pi\zeta}}\left\{ \cos\left(\zeta\mp\frac{\pi}{2}\left\vert \nu\right\vert -\frac{\pi}{4}\right)-\frac{1}{2\zeta}\left[\frac{\Gamma\left(\left\vert \nu\right\vert +\frac{3}{2}\right)}{\Gamma\left(\left\vert \nu\right\vert -\frac{1}{2}\right)}\right]\sin\left(\zeta\mp\frac{\pi}{2}\left\vert \nu\right\vert -\frac{\pi}{4}\right)\right\} \, ,\\
Y_{\pm\left\vert \nu\right\vert }\left(\zeta\right) & \simeq\sqrt{\frac{2}{\pi\zeta}}\left\{ \sin\left(\zeta\mp\frac{\pi}{2}\left\vert \nu\right\vert -\frac{\pi}{4}\right)+\frac{1}{2\zeta}\left[\frac{\Gamma\left(\left\vert \nu\right\vert +\frac{3}{2}\right)}{\Gamma\left(\left\vert \nu\right\vert -\frac{1}{2}\right)}\right]\cos\left(\zeta\mp\frac{\pi}{2}\left\vert \nu\right\vert -\frac{\pi}{4}\right)\right\} \, .
\end{align*} 
Because $\nu < 0$ and $w<0$, we obtain hyperbolic functions. In fact, using the properties 
of the Gamma functions (see e.g. Ref. \cite{Gradshteyn2007}), $\phi_{\bf{k}}$ is cast into the form:
\begin{align}
\phi_{\mathbf{k}}\left(\eta\right)= & \left(i\omega_{k}\eta\right)^{-\frac{3\epsilon}{\left(3\epsilon-2\right)}}\times \nonumber \\
 & \left\{ \left[\left(C_{{\bf{k}},1}+\frac{C_{{\bf{k}},2}}{\omega_{k}\eta}Q\right)X-\left(C_{{\bf{k}},2}-\frac{C_{{\bf{k}},1}}{\omega_{k}\eta}Q\right)Y\right]\cosh\left[\omega_{k}\left(\eta-\eta_{C}\right)\right]\right. \nonumber\\
 & \left.+\left[\left(C_{{\bf{k}},1}+\frac{C_{{\bf{k}},2}}{\omega_{k}\eta}Q\right)Z-\left(C_{{\bf{k}},2}-\frac{C_{{\bf{k}},1}}{\omega_{k}\eta}Q\right)W\right]\sinh\left[\omega_{k}\left(\eta-\eta_{C}\right)\right]\right\} \qquad  (k \eta \gg 1 ) \, , \label{eq:phikquasiL}
\end{align}
where
\[
\omega_{k}^{2}=(1-\epsilon)k^{2} \, ,
\]
and $\eta_C$ is the time of transition between the CDM and the DE phases. Quantities 
$\{Q,X,Y,Z,W\}$ are constants which depend on $\epsilon$ and $\eta_C$:
\begin{align}
U & = \frac{\pi}{2} \frac{3\epsilon}{\left( 3\epsilon-2 \right)}  \, , \nonumber \\
X & =\cos U\cosh\left(\omega_{k}\eta_{C}\right)+i\sin U\sinh\left(\omega_{k}\eta_{C}\right) \, , \nonumber \\
Y & =\sin U\cosh\left(\omega_{k}\eta_{C}\right)-i\cos U\sinh\left(\omega_{k}\eta_{C}\right)  \, , \nonumber \\
Z & =\cos U\sinh\left(\omega_{k}\eta_{C}\right)+i\sin U\cosh\left(\omega_{k}\eta_{C}\right)  \, , \nonumber \\
W & =\sin U\sinh\left(\omega_{k}\eta_{C}\right)-i\cos U\cosh\left(\omega_{k}\eta_{C}\right)  \, , \nonumber \\
Q & =\frac{1}{i} \frac{3\epsilon}{\left(3\epsilon-2\right)^{2}} \, . \label{Q}
\end{align}
With this, the potential $\phi_{\bf{k}}$ in the DE phase is determined up to the 
integration constants $C_{{\bf{k}},1}$ and $C_{{\bf{k}},2}$. These are
related to the constant $\bar{C}_{{\bf{k}},1}$ and $\bar{C}_{{\bf{k}},2}$ in the matter phase
solution by general relativistic matching conditions. This is the subject of the next subsection.


\subsection{Matching Conditions \label{sec:DMconditions}}

In our model there is an instantaneous transition in the equation of state
at redshift $z_C$. On either side of the matching surface the Einstein
equations are satisfied. Across the surface, the metric must obey matching
conditions which are a generalization of the Israel matching conditions \cite{Israel}
and which were derived in  Ref. \cite{Deruelle1995} (see also \cite{Hwang}).

Let us use the index $(-)$ as a label for quantities in the CDM regime and $(+)$ 
for quantities in the DE regime. The CDM fluctuations $\phi_{{\bf{k}}-}$ and $\delta_{{\bf{k}}-}$ 
are then connected to the DE ones, $\phi_{{\bf{k}}+}$ and $\delta_{{\bf{k}}+}$, through 
the matching conditions
\begin{equation}
\phi_{{\bf{k}}-}(\eta_C) = \phi_{{\bf{k}}+}(\eta_C) \qquad \text{and} \qquad \frac{\delta_{{\bf{k}}-}(\eta_C)}{(1+w_{-})} = \frac{\delta_{{\bf{k}}+}(\eta_C)}{(1+w_{+})}  \, .\label{eq:DMconditions}
\end{equation}
Inserting the sub-Hubble solutions in the CDM and DE periods derived in the
previous subsections we obtain \footnote{Making use of $w_{-} = w_{\text{CDM}} = 0$ 
and $w_{+} = w_{\text{DE}} = ( -1 + \epsilon )$.}:
\begin{eqnarray}
\phi_{{\bf{k}}-}\left(\eta_{C}\right) &=& \frac{1}{\left(i\omega_{k}\eta_{C}\right)^{\frac{3\epsilon}{\left(3\epsilon-2\right)}}} \left[\left(C_{{\bf{k}},1}+\frac{C_{{\bf{k}},2}Q}{\omega_{k}\eta_{C}}\right)X-\left(C_{{\bf{k}},2}-\frac{C_{{\bf{k}},1}Q}{\omega_{k}\eta_{C}}\right)Y\right] \, ;
\nonumber\\
\phi_{\mathbf{k}-}\left(\eta_{C}\right) &\times& \left\{ -\frac{1}{6}\left(k\eta_{C}\right)^{2} 
\left[\left(-1+\epsilon\right)\left(3\epsilon-2\right)\right]-\frac{3\epsilon}{\left(3\epsilon-2\right)}\right\} \label{eq:system-LS} \\
&=&\frac{\left(\omega_{k}\eta_{C}\right)}{\left(i\omega_{k}\eta_{C}\right)^{\frac{3\epsilon}{\left(3\epsilon-2\right)}}}\left[\frac{Q}{\left(\omega_{k}\eta_{C}\right)^{2}}\left(C_{{\bf{k}},2}X+C_{{\bf{k}},1}Y\right)-\left(C_{{\bf{k}},1}+\frac{C_{{\bf{k}},2}}{\omega_{k}\eta_{C}}Q\right)Z+\left(C_{{\bf{k}},2}-\frac{C_{{\bf{k}},1}}{\omega_{k}\eta_{C}}Q\right)W\right] \, ,
\nonumber
\end{eqnarray}
where 
\begin{equation}
\phi_{{\bf{k}}-}\left(\eta_{C}\right) = \bar{C}_{{\bf{k}},1} + \frac{ \bar{C}_{{\bf{k}},2} }{\left[ (3\epsilon - 2) \eta_{C} \right]^{5} } \, . \label{eq:phi_minus(etaC)}
\end{equation}

The factor $(3\epsilon - 2)$ in Eq.~(\ref{eq:phi_minus(etaC)}) comes from the 
integration constants we neglected when writing Eq.~(\ref{eq:calH(eta)wconstant}) 
for $\mathcal{H} \propto \eta^{-1}$. The details about these constants can be found in the 
Appendix. 

The system in Eq.~(\ref{eq:system-LS}) can be solved to find the constants 
$\{C_{{\bf{k}},1},C_{{\bf{k}},2}\}$ appearing in $\phi_{{\bf{k}}+}$ in terms of 
the constants $\{\bar{C}_{{\bf{k}},1},\bar{C}_{{\bf{k}},2}\}$ of the CDM phase. They,
in turn, are determined in terms of the primordial spectrum of fluctuations and
the evolution until the time of equal matter and radiation.  Note that even if 
$\bar{C}_{{\bf{k}},2}$ is of the same magnitude of $\bar{C}_{{\bf{k}},1}$ at the 
end of the phase of very early universe cosmology when the fluctuations
are generated, it is the coefficient of the decaying mode of $\phi_{\bf{k}}$ which 
can be neglected. 

In solving for  $\{C_{{\bf{k}}1},C_{{\bf{k}},2}\}$ we keep only terms scaling with 
the highest power of $(\omega_k \eta_C)$ (this is consistent in the sub-Hubble case). 
The constants in Eq.~(\ref{eq:phikquasiL}) are then determined to be:
\begin{eqnarray}
C_{{\bf{k}},1} &=& \frac{\left(i\omega_{k}\eta_{C}\right)^{\frac{3\epsilon}{\left(3\epsilon-2\right)}}\phi_{-}\left(\eta_{C}\right)}{\left[ XW-YZ \right]} 
\left\{  \frac{1}{6} \left( 3\epsilon-2 \right)  \left[ \left(\omega_{k}\eta_{C}\right) Y - QX \right] + W \right\} \\
C_{{\bf{k}},2} &=& \frac{\left(i\omega_{k}\eta_{C}\right)^{\frac{3\epsilon}{\left(3\epsilon-2\right)}}\phi_{-}\left(\eta_{C}\right)}{\left[ XW-YZ \right]}
\left\{ \frac{1}{6} \left( 3\epsilon - 2 \right) \left[ \left(\omega_{k}\eta_{C}\right) X + QY \right] + Z \right\}
\nonumber
\end{eqnarray}
Then, by using the identities
\[
\left[X^{2}+Y^{2}\right]=1 \, , \qquad \left[XZ+YW\right]=0  \, , \qquad \left[WX-YZ \right]=-i \, ,
\]
we, at last, get $\phi_{+}$ in its final form:
\begin{align}
\phi_{\mathbf{k}+}\left(\eta\right) & =\phi_{\mathbf{k}-}\left(\eta_{C}\right)\left(\frac{\eta_{C}}{\eta}\right)^{\frac{3\epsilon}{\left(3\epsilon-2\right)}}\left\{ \left[1-\frac{1}{6}\frac{3\epsilon}{\left(3\epsilon-2\right)}\left(1-\frac{\eta_{C}}{\eta}\right)\right]\cosh\left[\omega_{k}\left(\eta-\eta_{C}\right)\right]\right.\nonumber \\
 & \qquad\qquad\qquad\qquad\qquad\quad\left.+\left[-\frac{1}{6}\left(3\epsilon-2\right)\left(\omega_{k}\eta_{C}\right)\right]\sinh\left[\omega_{k}\left(\eta-\eta_{C}\right)\right]\right\} \,.\label{eq:phiplus}
\end{align}
Notice that the expression for the DE-regime $\phi_{{\bf{k}}+}$ reduces to the 
result in the CDM-phase $\phi_{{\bf{k}}-}$ at $\eta = \eta_C$ as it must
according to the matching conditions. 

Note that the 
$k$-dependence of $\phi_{{\bf{k}}+}$ appears both in the argument of the hyperbolic 
functions and in the coefficient of $\sinh$ through $\omega_k = \sqrt{1-\epsilon} \, k$. 
Eq.~(\ref{eq:phiplus}) for $\phi_{{\bf{k}}+}$ can be substituted into 
Eq.~(\ref{eq:densitycontrast(phi)}) in order to obtain $\delta_{{\bf{k}}+}$. The expression is 
long. Keeping only the dominant terms in powers of $\left(k\eta\right)$,
it reduces to: 
\begin{align}
\delta_{\mathbf{k}+}\left(\eta\right) & \simeq\delta_{\mathbf{k}-}\left(\eta_{C}\right)\left(\frac{\eta_{C}}{\eta}\right)^{-\frac{\left(3\epsilon-4\right)}{\left(3\epsilon-2\right)}}\left\{ \left[\epsilon\left(\frac{\eta_{C}}{\eta}\right)+\left[1-\frac{1}{6}\frac{3\epsilon}{\left(3\epsilon-2\right)}\right]\left(1-\frac{\eta_{C}}{\eta}\right)\right]\cosh\left[\omega_{k}\left(\eta-\eta_{C}\right)\right]\right.\nonumber \\
 & \qquad\qquad\qquad\qquad\qquad\quad\left.+\left[-\frac{1}{6}\left(3\epsilon-2\right)\left(\omega_{k}\eta_{C}\right)\right]\sinh\left[\omega_{k}\left(\eta-\eta_{C}\right)\right]\right\} \label{eq:deltaplus}
\end{align}
where 
\begin{equation}
\delta_{\mathbf{k}-}\left(\eta_{C}\right)\simeq\left[-\frac{1}{6}\left(3\epsilon-2\right)^{2}\left(k\eta_{C}\right)^{2}\right]\phi_{\mathbf{k}-}\left(\eta_{C}\right)\,.\label{eq:deltaminus}
\end{equation}
Notice that Eq.~(\ref{eq:deltaplus}) guarantees that
\begin{equation}
\delta_{\mathbf{k}+}\left(\eta_{C}\right)=\epsilon\delta_{\mathbf{k}-}\left(\eta_{C}\right)
\end{equation}
as required by Eq.~(\ref{eq:DMconditions}).

We will apply these matching conditions in two different cases. First, we shall consider 
modes which enter the Hubble radius after the time of equal matter and radiation. We call
these the ``large-scale'' fluctuations (Sect. \ref{sec:DM-LS}). Then, in Sect. \ref{sec:DM-SS}, 
we take consider ``small-scale'' perturbations which enter the Hubble radius in the
radiation phase \footnote{Note that we consider only fluctuations which are sub-horizon 
at the transition redshift $z_C$.}. In both cases we will be matching sub-Hubble modes
at the transition surface $z = z_C$. The difference between these two cases is the
evolution of the fluctuations once they enter the Hubble radius. For large-scale
fluctuations the primordial spectrum is unchanged, for small-scale fluctuations the
primordial spectrum is processed since the fluctuation modes grow only logarithmically
between when they enter the Hubble radius and the time of equal matter and radiation.

\subsubsection{Large-scale fluctuations \label{sec:DM-LS}}

Large scales enter the Hubble radius when the universe is already dominated by 
matter (but before $z_C$).  Therefore, the coefficients in Eqs.~(\ref{eq:phikdust}) 
and (\ref{eq:deltakdustsub}) for $\phi_{{\bf{k}}-}$ and $\delta_{{\bf{k}}-}$ 
are determined directly by the primordial spectrum which needs to be
nearly scale-invariant to agree with observations. Accordingly, we introduce the notation $\phi_{{\bf{k}}-} = \bar{C}_{{\bf{k}},1} = \phi_{{\bf{k}}\text{p}}^{\text{L}}$ where label ``L'' stands for large-scale and ``p'' for primordial.

Note that for a scale-invariant primordial spectrum of curvature fluctuations such
as emerges from an early phase of inflation \cite{Mukh} (or alternatives such
as the ``matter bounce'' scenario \cite{Fabio} or ``string gas cosmology'' \cite{BV, NBV})
\begin{equation}
\phi_{\mathbf{k}-}  \, \sim \, k^{-3/2} \, ,
\end{equation} 
and hence
\begin{equation}
\delta_{\mathbf{k}-} (\eta) \, \sim \, k^{1/2} \, .
\end{equation} 

Then, Eq. (\ref{eq:phikquasiL}) for $\phi_{{\bf{k}}+}$ can be used to calculate $\delta_{{\bf{k}}+}$ through 
the ``Poisson'' Eq.~(\ref{eq:densitycontrast(phi)}). 

\subsubsection{Small-scale fluctuations \label{sec:DM-SS}}

Small scale modes enter the Hubble radius when the universe is radiation dominated. 
The radiation pressure will only allow the density fluctuation modes to grow logarithmically
in time between when they enter the Hubble radius and the time of equal matter
and radiation. This is the Meszaros effect \cite{Meszaros1974}. This means
that $\delta_{\bf{k}}$ will evolve as 
\begin{equation}
\delta_{\bf{k}} \sim \ln (k \eta) 
\end{equation}
after the mode enter the horizon (see e.g. \cite{Dodelson2003}). After the time
of equal matter and radiation regime the fractional density contrast will then grow
linearly in the scale factor, i.e. proportional to $\sim a/a_{\text{eq}}$ 
\footnote{Quantities with subscript $eq$ are calculated at the time $\eta_{\text{eq}}$
of equal  matter and radiation energy density.}. Hence
\begin{equation}
\delta_{{\bf{k}}-}\left(\eta\right)=K_{{\bf{k}}} \left( \frac{a}{a_\text{eq}} + \frac{2}{3}\right) \simeq K_{{\bf{k}}} \left[ \frac{1}{4} H_0^2 \Omega_{\text{DE},0} \left( \frac{a_C}{a_\text{eq}} \right) \left( \frac{a_0}{a_C} \right)^{\left( 3\epsilon - 2 \right)} \right]  \left( \eta + 3 w_{+} \eta_C \right)^{2}\,,\label{eq:deltaminusSS}
\end{equation}
if $a\gg a_\text{eq}$ where $w_{+} = -1+\epsilon$ and
\begin{equation}
K_{{\bf{k}}}=\frac{3}{2}\left(A\phi_{{\bf{k}}\text{p}}^{\text{S}}\right)\ln\left[ \frac{4B}{e^{3}}  \sqrt{2} \frac{k}{k_{\text{eq}}}\right]\, .\label{eq:K}
\end{equation}
Here, the constant $\phi_{{\bf{k}}\text{p}}^{\text{S}}$ (label ``S'' means ``small scales'')
comes from the primordial spectrum, $A=9.0$ and $B=0.62$ are numbers given by 
Dodelson in \cite{Dodelson2003} (and are not affected by the DE phase of our model), 
$\Omega_{\text{DE},0}$ is the density parameter of the dark fluid today, and $a_C = a(\eta_C)$ 
is the scale factor at CDM-DE transition from which we build 
$z_C = \left(a_0/ a_C\right) -1$. Finally, by following the procedure explained in 
Ref.~\cite{Dodelson2003}, the scale $k_{\text{eq}}$ (the comoving wavenumber
entering the Hubble radius at $t_{eq}$) is calculated to be:
\begin{equation}
k_{\text{eq}} = \sqrt{2} H_0 \frac{\Omega_{\text{DE},0}}{\sqrt{\Omega_{\gamma 0}}}  \left( 1 + z_C \right)^{3 w_{+}}\, . \label{eq:keq}
\end{equation}
This equation contains $\Omega_{\gamma0}$, the present-day value of radiation 
density parameter. Eq.~(\ref{eq:deltaminusSS}) uses the background solution 
(\ref{eq:awconstant}) with the integration constants obtained from continuity of 
$a(\eta)$ and $\cal{H}(\eta)$ at $\eta=\eta_{\text{eq}}$. The interested reader is 
referred to the Appendix where the details are worked out.

By reverse engineering involving the ``Poisson'' Eq.~(\ref{eq:densitycontrast(phi)}) and 
Eq. (\ref{eq:deltaminusSS}), we determine:
\begin{equation}
\phi_{{\bf{k}}-} \simeq - K_{{\bf{k}}} \left[ \frac{3}{2} H_0^2 \Omega_{\text{DE},0} \left( \frac{a_C}{a_\text{eq}} \right) \left( \frac{a_0}{a_C} \right)^{\left( 3\epsilon - 2 \right)} \right] \frac{1}{k^2} \,.\label{eq:phiminusSS}
\end{equation}

With the small-scale $\phi_{{\bf{k}}-}(\eta)$ and  $\delta_{{\bf{k}}-}(\eta)$ at hand --
Eqs.~(\ref{eq:deltaminusSS}) and (\ref{eq:phiminusSS}), the next step is to obtain 
$\phi_{{\bf{k}}+}(\eta)$ and  $\delta_{{\bf{k}}+}(\eta)$ good for small scales. 
As small-scale DE fluctuations also evolve in a DE phase for the background, all the 
reasoning involving $\phi_{{\bf{k}}+}$ in Sect. \ref{sec:DE-Perturb} -- previous to the 
imposition of the matching conditions in Sect. \ref{sec:DM-LS} -- still holds; i.e. 
Eq.~(\ref{eq:phikquasiL}) remains unchanged and also the related $\delta_{{\bf{k}}+}$. 
This $\delta_{{\bf{k}}+}$ is then glued to $\delta_{-}$, Eq. (\ref{eq:deltaminusSS}), 
according to the matching conditions. Also, $\phi_{{\bf{k}}+}$ as given by 
Eq. (\ref{eq:phikquasiL}) is set equal to $\phi_{{\bf{k}}-}$, Eq. (\ref{eq:phiminusSS}), 
at $\eta=\eta_C$. This way, constants $\left\{ C_{{\bf{k}},1},C_{{\bf{k}},2} \right\}$ for 
small scales are calculated.

\subsubsection{Summary of the Results}

The result for the small-scale $\phi_{{\bf{k}}+}$ is formally the same as the 
large-scales $\phi_{{\bf{k}}+}$, but because the functional form of $\phi_{{\bf{k}}-}$ changes, 
the $k$-dependence is different in each case. In fact, 
\begin{align}
\phi_{{\bf k}+}^{\text{L}}\left(\eta\right) & =\phi_{{\bf k}\text{p}}^{\text{L}}\left(\frac{\eta_{C}}{\eta}\right)^{\frac{3\epsilon}{\left(3\epsilon-2\right)}}\left\{ \left[1-\frac{1}{6}\frac{3\epsilon}{\left(3\epsilon-2\right)}\left(1-\frac{\eta_{C}}{\eta}\right)\right]\cosh\left[\sqrt{1-\epsilon}\left(k\eta-k\eta_{C}\right)\right]\right.\nonumber \\
 & \qquad\qquad\qquad\qquad\quad\left.+\left[-\frac{1}{6}\left(3\epsilon-2\right)\sqrt{1-\epsilon}\left(k\eta_{C}\right)\right]\sinh\left[\sqrt{1-\epsilon}\left(k\eta-k\eta_{C}\right)\right]\right\} \label{eq:phikplusLS}
\end{align}
(the superscript ``L'' standing for large-scale) while
\begin{align}
\phi_{\mathbf{k}+}^{\text{S}}\left(\eta\right) & =\left[-\left(\frac{3}{2}\right)^{2}H_{0}^{2}\Omega_{\text{DE},0}\left(\frac{a_{C}}{a_{\text{eq}}}\right)\left(\frac{a_{0}}{a_{C}}\right)^{\left(3\epsilon-2\right)}A\right]k^{-2}\ln\left[\frac{4B}{e^{3}}\sqrt{2}\frac{k}{k_{\text{eq}}}\right]\times\nonumber \\
 & \phi_{\mathbf{k}\text{p}}^{\text{S}}\left(\frac{\eta_{C}}{\eta}\right)^{\frac{3\epsilon}{\left(3\epsilon-2\right)}}\left\{ \left[1-\frac{1}{6}\frac{3\epsilon}{\left(3\epsilon-2\right)}\left(1-\frac{\eta_{C}}{\eta}\right)\right]\cosh\left[\sqrt{1-\epsilon}\left(k\eta-k\eta_{C}\right)\right]\right.\nonumber \\
 & \qquad\qquad\qquad\quad\left.+\left[-\frac{1}{6}\left(3\epsilon-2\right)\sqrt{1-\epsilon}\left(k\eta_{C}\right)\right]\sinh\left[\sqrt{1-\epsilon}\left(k\eta-k\eta_{C}\right)\right]\right\} \,.\label{eq:phikplusSS}
\end{align}
This has a direct impact on the power spectrum \cite{Sparke2007} predicted by these two regimes, 
as we shall see in the next section.




\section{Unified dark fluid power spectrum \label{sec:PS}}

The current power spectrum of density fluctuations can be defined in terms of $\delta_{\bf{k}}(\eta)$ as:
\begin{equation}
P_{{\bf k}}\left(\eta \right) = 2\pi^2 \left|\delta_{{\bf k}}\left(\eta \right)\right|^{2}\label{eq:Pk}
\end{equation}
where $P_{{\bf k}}$ must be calculated today $(\eta = \eta_0)$ for comparison with observations \cite{Cole2005,Sanchez2006}. The observational power spectrum is a rough measure of 
the distribution of matter. In our unified model, cold dark matter evolves into a dark
energy fluid. Hence, it is the current dark energy power spectrum that would have to coincide 
with the observed matter power spectrum today. 

In order to agree with observations, we need to obtain the following 
asymptotic limits on large and small scales, respectively \cite{Dodelson2003}:
\begin{equation}
P_{\bf{k}}^{\text{L}} \left( \eta_0 \right) \sim k \qquad (k \ll k_{\text{eq}}) \label{eq:Pk_LS_Data}
\end{equation} 
and
\begin{equation}
P_{\bf{k}}^{\text{S}} \left( \eta_0 \right) \sim k^{-3} \left( \ln k \right)^2 \qquad (k \gg k_{\text{eq}}) \, . \label{eq:Pk_SS_Data}
\end{equation}

Our model must give a power spectrum which agrees with $P_{\bf{k}}$  from (\ref{eq:Pk_LS_Data}) 
and (\ref{eq:Pk_SS_Data}) within the observational error bars to be viable. However, we 
have seen that a factor of $\cosh \left[ \omega_k \left( \eta - \eta_C \right) \right]$ appears in $\phi_{{\bf{k}}+} $ for both large and small scales -- Eqs. (\ref{eq:phikplusLS}) and (\ref{eq:phikplusSS}). 
This factor introduces instabilities which grow both in $\eta$ and in $k$, and
these instabilities propagate to $\delta_{{\bf{k}}+}$ and consequently to $P_{\bf{k}}$. 

In fact, by using the background evolution $a(\eta)$, it is possible to express 
$P_{\bf{k}}(\eta_0)$ for the dark fluid in the general form:
\begin{equation}
P_{\bf{k}}(\eta_0) \sim (\phi_{\bf{k}\text{p}})^2  \times \{\text{usual }k\text{-behavior}  \} \times \{\text{Enhancement Factor }\mathcal{F}(k,z_C,\epsilon,\Omega_{\text{DE},0})\}  \label{eq:Pk_general_form}
\end{equation}
which is valid both for large scales and small scales. Here, the first factor on the right hand side,
$\phi_{\bf{k} \text{p}} \propto k^{-3/2}$, 
is the primeval spectrum of the potential \cite{Dodelson2003} (assuming an initial
scale-invariant spectrum). It is related to the curvature perturbation $\mathcal{R}$ on 
super-Hubble scales via \cite{Brandenberger2004}:
\begin{equation}
\mathcal{R} = \left[ 1 + \frac{2}{3} \frac{1}{(1 + w)} \right] \phi_{\text{p}} \, . \label{calR}
\end{equation}
Large-scale fluctuations enter the horizon in a matter dominated universe, when $w = 0$. Conversely, small-scale perturbations make the transition from super-horizon to sub-horizon in a radiation dominated universe with $w = 1/3$. Therefore, Eq. (\ref{calR}) gives:
\begin{equation}
{\cal R}=\begin{cases}
\frac{5}{3}\phi_{\text{p}}^{\text{L}} & \qquad\left(w=0\right)\\
\frac{3}{2}\phi_{\text{p}}^{\text{S}} & \qquad\left(w=1/3\right)
\end{cases} \, . \label{As}
\end{equation}
The second factor on the right hand side of (\ref{eq:Pk_general_form}) is the usual
transfer function of density fluctuations which takes into account the different evolution
of large and small scale density fluctuations (discussed at the end of the previous
section), and the final factor is the extra growth factor in our model compared to what
is obtained in the $\Lambda$CDM scenario.

The specific functional form of Eq.~(\ref{eq:Pk_general_form}) on large scales is
\begin{equation}
P_{{\bf k}}^{\text{L}}\left(\eta_{0}\right)  \simeq  \left( \Phi_{\bf{k}\text{p}}^{\text{L}} \right)^2 \left\{ \frac{c}{H_0} k \right\} \mathcal{F}\left( k,z_C,\epsilon,\Omega_{\text{DE},0} \right)  \, , \label{eq:Pklarge}
\end{equation}
with
\begin{equation}
\Phi_{\bf{k}\text{p}}^{\text{L}}  \equiv \sqrt{2 \pi^2} \left( - \frac{2}{3}  \phi_{\bf{k}\text{p}}^{\text{L}} \right) 
\left( \frac{c}{H_0} \right)^{\frac{3}{2}}  \label{eq:PhipLS}
\end{equation}
and the enhancement factor
\begin{equation}
\mathcal{F} = \mathcal{F}\left( k,z_C,\epsilon,\Omega_{\text{DE},0} \right)  \equiv \left\{ \frac{\left( 1 + z_C \right)^{ \left(3w_{+}+2\right)}}{\Omega_{\text{DE},0}} \left[ \left( \epsilon + \zeta_C \right) \cosh \frac{k \tilde{\eta}_C}{\sqrt{\Omega_{\text{DE},0}}}   - \frac{1}{3} \frac{c}{H_0}\frac{\sqrt{1-\epsilon} k}{\sqrt{\Omega_{\text{DE},0}}}  \sinh \frac{k \tilde{\eta}_C}{\sqrt{\Omega_{\text{DE},0}}}  \right] \right\}^2 \, , \label{eq:calF}
\end{equation}
where we have defined
\begin{equation}
\tilde{\eta}_C = \tilde{\eta}_C \left( z_C,\epsilon \right) \equiv  \left( \frac{c}{H_0}  \right) 2 \frac{\sqrt{1-\epsilon}}{\left( 3\epsilon -2 \right)} \left[ 1 - \left( 1 + z_C \right)^{\frac{\left( 3\epsilon - 2 \right)}{2}} \right] \label{eq:eta_tilde_C}
\end{equation}
and
\begin{equation}
\zeta_C = \zeta_C \left( z_C,\epsilon \right) \equiv \left[ 1 - \frac{1}{2} \frac{\epsilon}{ \left( 3\epsilon - 2 \right)} \right] \left[ \left( 1 + z_C \right)^{- \frac{\left( 3\epsilon - 2 \right)}{2}} - 1 \right]\, . \label{eq:zeta_C}
\end{equation}
Note that the enhancement factor introduces a departure from the shape of the
initial spectrum which grows exponentially in $k$. Moreover,
\[
\lim_{z_C \rightarrow 0} \tilde{\eta}_C = \lim_{z_C \rightarrow 0} \zeta_C = 0 \, . 
\]
Eq. (\ref{eq:Pklarge}) for $P_{{\bf k}}^{\text{L}}\left(\eta_{0}\right) $ holds as long as $k \ll k_{\text{eq}}$.

The small scale (S) power spectrum for the dark fluid today is:
\begin{equation}
P_{{\bf k}}^{\text{S}}\left(\eta_{0}\right) \simeq \left( \Phi_{\bf{k}\text{p}}^{\text{S}} \right)^2  \left\{ \left( \frac{k_{\text{eq}}}{\sqrt{2}k} \right)^3 \ln^2 \left[ \frac{4B}{e^3} \frac{\sqrt{2} k}{k_{\text{eq}}} \right]  \right\} \mathcal{G}\left( k,z_C,\epsilon,\Omega_{\text{DE},0} \right) \, , \label{eq:Pksmall}
\end{equation}
with
\begin{equation}
\Phi_{\text{p}}^{\text{S}} \equiv \sqrt{2 \pi^2} \left( \frac{3}{2} A  \phi_{\text{p}}^{\text{S}}  \right) 
\left( \frac{c}{H_0} \right)^{\frac{3}{2}} \, , \label{eq:PhipSS}
\end{equation}
and
\begin{equation}
\mathcal{G}\left( k,z_C,\epsilon,\Omega_{\text{DE},0} \right) \equiv \frac{\Omega_{\text{DE},0}}{\sqrt{\Omega_{\gamma0}}} \left( 1 + z_C \right)^{3w_{+}} \mathcal{F}\left( k,z_C,\epsilon,\Omega_{\text{DE},0} \right) \label{eq:calG}
\end{equation}
given in terms of the enhancement factor $\mathcal{F}$ in Eq.~(\ref{eq:calF}). 
Eq. (\ref{eq:Pksmall}) for $P_{{\bf k}}^{\text{S}}\left(\eta_{0}\right) $ holds as 
long as $k \gg k_{\text{eq}}$. The enhancement factor once again introduces an exponential
divergence from the shape of the initial spectrum.

\begin{figure}[h!]
\begin{center}
\includegraphics[width=0.525\textwidth]{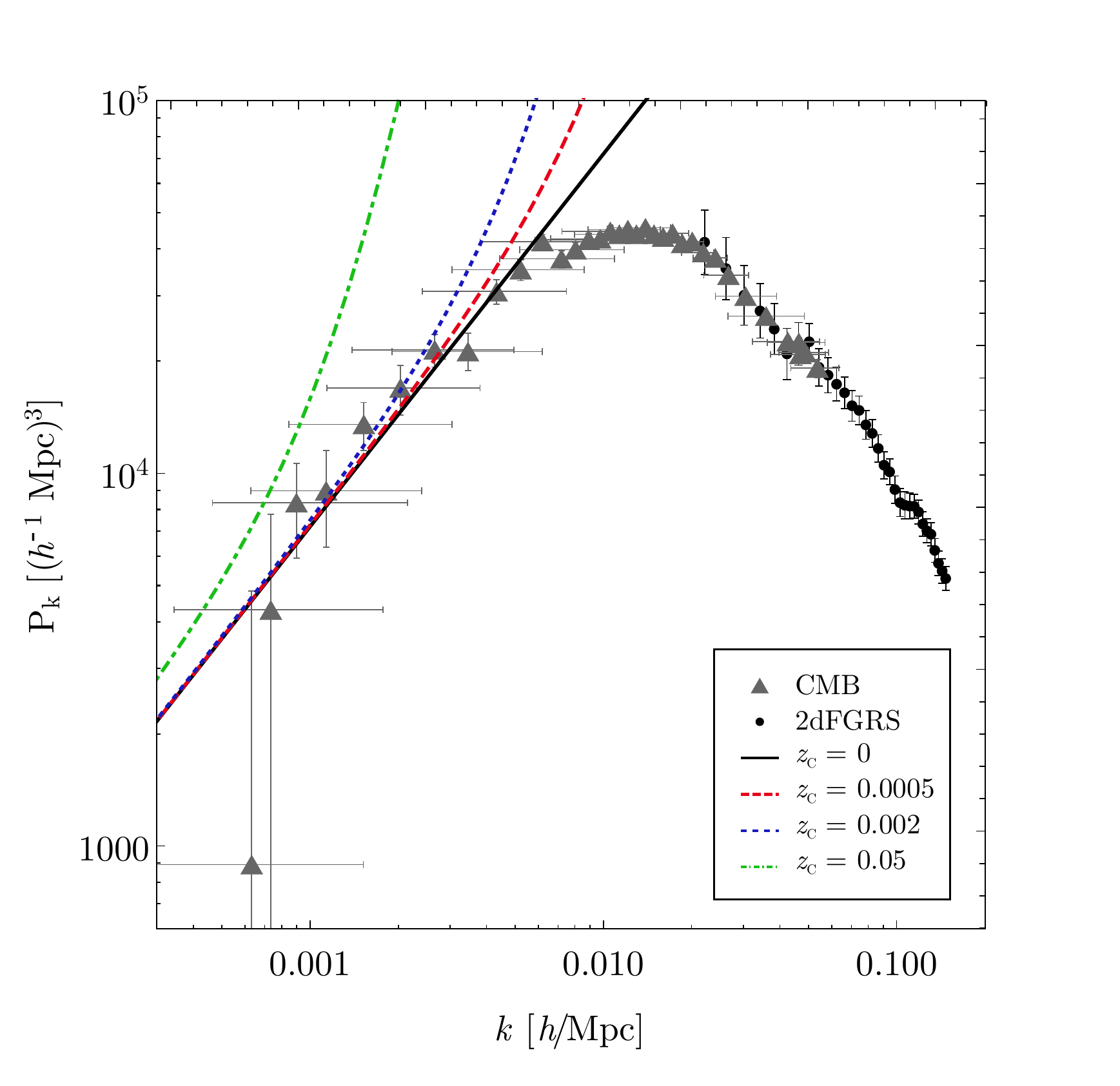}
\includegraphics[width=0.525\textwidth]{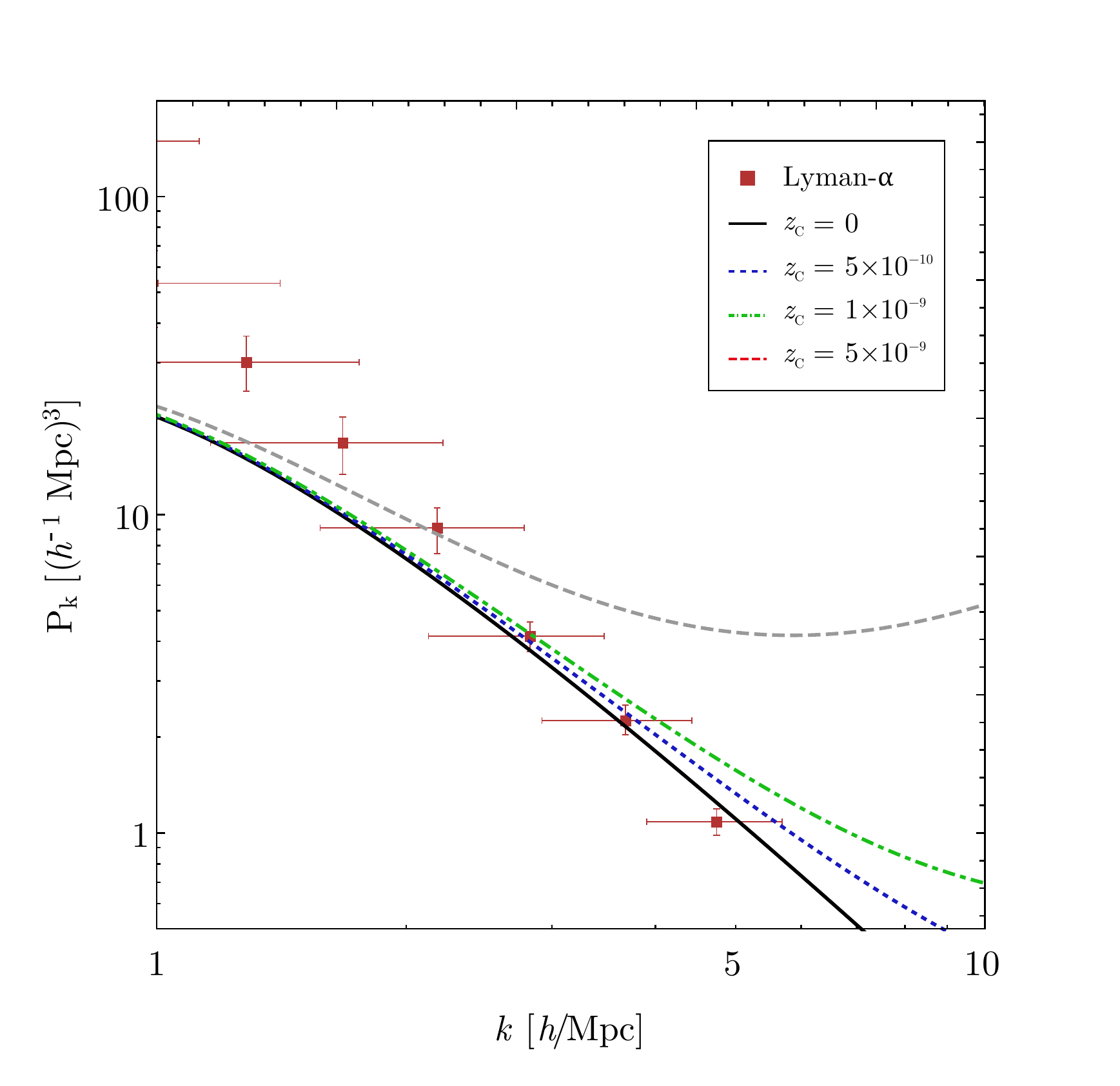}
\end{center}
\caption{Observational and predicted values of $P_{\bf{k}}$ for  
our unified dark matter model with piecewise
constant equation of state. Both large-scale fit (upper plot) and small-scale fit use 
$\epsilon = 0.28$ and $\Omega_{\text{DE},0} = 1$. The justification for this particular 
value of $\epsilon$ is given in Sect. \ref{sec:Intro}. We take the present-day value of the 
dark component to be one because we neglect the 4\% contribution coming from baryonic
matter to simplify the treatment.}
\label{Fig:Pk_LS_SS}
\end{figure}

\begin{figure}[h!]
\begin{center}
\includegraphics[width=0.5\textwidth]{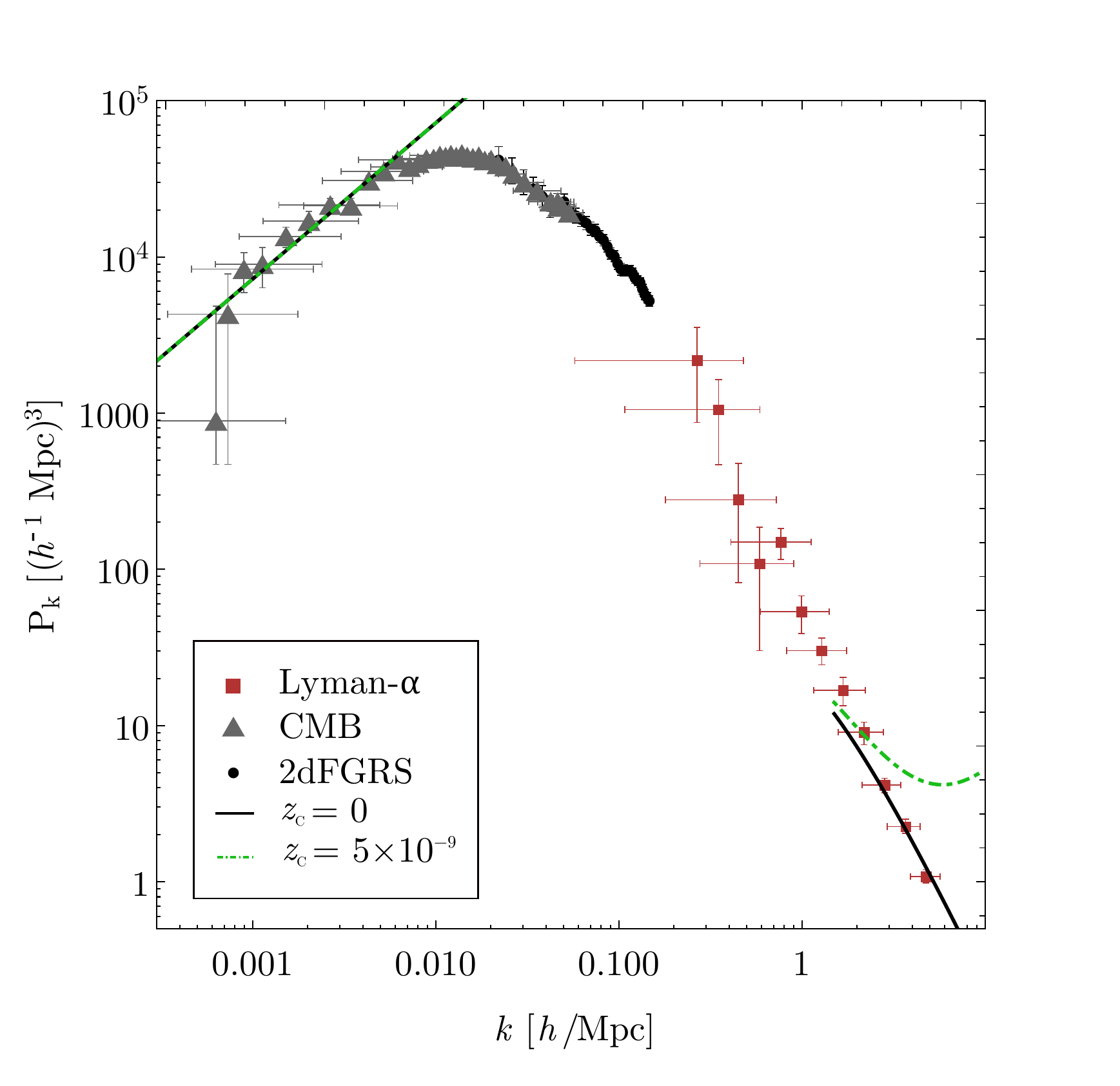}
\end{center}
\caption{$P_{\bf{k}}(\eta_0)$ for the unified dark sector with a sharp transition. 
Large-scale curves are valid for $k \ll k_{\text{eq}}$ when the data points come
from CMB and large-scale structure data. The small-scale curves are compared to 
Lyman-Alpha Forest data for $k \gg k_{\text{eq}}$. Small values of $z_C$ which are
consistent with large-scale points already causes a strong divergence at small scales.}
\label{Fig:Pk}
\end{figure}

We perform a numerical fit of the power spectrum to the data which indicates that  
exponential divergence $P_{{\bf k}}(\eta_0) $ (see (\ref{eq:Pklarge}) and (\ref{eq:Pksmall})) 
caused by the enhancement factor (\ref{eq:calF}) can only be reconciled with the
data if the transition redshift $z_C$ is extremely close to $z_C = 0$ -- see the 
plots in Figs.~\ref{Fig:Pk_LS_SS} and \ref{Fig:Pk}. On large scales, the observational
data come from CMB anisotropies (see the points labelled CMB in the plots) and 
from the power spectrum of galaxies on large scales (see the points
labelled 2dFGRS). On small scales the tightest observational constraints come from Lyman alpha observations (the Lyman-$\alpha$ points on the graphs) \footnote{Note that there are a number of references that use the Lyman-alpha Forest in the linear regime, see e.g. \cite{McDonald2005,McDonald2007,Seljak2006}.}.


In these figures, we have independently fixed the normalization of the primordial
power spectrum to either obtain agreement with the data in the $k \rightarrow 0$
limit (in the case of large scales) or at $k \sim 1 \times h {\rm{Mpc}}^{-1}$ (in the
case of small scales). In fact, we should use the small-scale normalization also
on larger scales. We see that the tightest constraints on the redshift $z_C$ come
from the smallest scales which are in the linear regime. With values of $z_C$ for
which the small-scale power spectrum is consistent with the data, the large-scale
spectrum is also consistent.


The only hope for our unified model is to take the transition time $\eta_C$ 
to be very close to the present-day time $\eta_0$. In this limit 
$\cosh \left[ \omega_k \left( \eta - \eta_C \right) \right] \rightarrow 1$ and 
$\sinh \left[ \omega_k \left( \eta - \eta_C \right) \right] \rightarrow 0$.




\section{Discussion\label{sec:Conclusions}}

We studied cosmological perturbations in a unified dark fluid model in which perfect
fluid matter with an initial cold dark matter equation of state
$w=0$  evolves to a DE regime characterized by a constant 
EoS parameter of the type $w = -1 + \epsilon$ (cf. Sect. \ref{sec:Intro}) and the
transition is assumed to be instantaneous. In this simple model the equations for the relativistic 
potential $\phi_{\bf{k}}$ can be solved analytically (Sects. \ref{sec:PerturbEqs} and 
\ref{sec:PerturbUM}). The solutions on either side of the transition hypersurface 
must be connected using the general relativistic matching conditions. The power 
spectrum (PS) of density fluctuations was computed both in the limit of large scales 
and small scales (Sect. \ref{sec:PS}). Due to the negative speed of sound in
the dark energy phase which is a consequence of the perfect fluid description of
our matter, there is an exponential instability both in time and in comoving wavenumber $k$ which leads both to an exponential growth of the amplitude of the power spectrum and to a shape distortion. Demanding agreement with the observed spectrum requires setting the transition redshift between the two phases extremely close to zero, $z_C = 0$ \footnote{Note that the constraint on the transition redshift $z_C$ would be even tighter had we taken into account the nonlinear effects which boost the power spectrum on small scales, in particular on the QSO scales. We have provided a conservative bound, but this bound already rules out the model.}. If we do this, however, the model is unable to match
the background cosmology data \cite{Planck2015,BOSS,GRB,SNIa}. This 
rules out a unified dark sector model given by perfect fluid matter with a piecewise
constant equation of state. Our analytical approach reveals the origin of 
the instabilities faced by some unified dark sector models. Our analytical
approach also makes it clear that the problems cannot be solved by smoothing
the transition.

We have seen in Sect. \ref{sec:Intro} that our model can be obtained as 
particular limits of other UDM models at the background level. The general 
results of our analysis may be applicable to these other models. In particular, this is  
the case for GCG in the range $\alpha < 0$ -- see Eq. (\ref{eq:w_GCG}). 
Ref. \cite{Sandvik2004}  showed that pronounced divergences in the GCG matter power 
spectrum appear for $| \alpha | > 10^{-5}$. Our analysis enables us to pinpoint the 
origin of these divergences in the power spectrum: they come about early in the
DE-phase due to the exponential increase (both in time and in $k$) of the relativistic 
potential.

Another way to construct unified models of dark matter and dark energy is to make
use of scalar fields instead of perfect fluids. In this case $c_s^2 \neq w$
(see e.g. the review in \cite{Bertacca2010}).  In particular, $c_s^2$ can be positive
semi-definite, and hence the instability of fluctuations which we have encountered
in our dark fluid model does not arise. For canonical scalar fields $c_s^2 = 1$.
This then gives rise to an opposite problem: at the level of linear cosmological
perturbations it becomes impossible to explain the observed gravitational
clustering of dark matter. This problem can be avoided by ceasing to regard
the matter field to be coherent \footnote{This is how axion dark matter works.}. 
For recent examples of unified scalar field models of dark matter and
dark energy see \cite{Juerg, Stephon, Elisa}.

In spite of the limits in which our unified model with piecewise constant equation
of state can be obtained from other unified models (such as GCG, the arctan-UDM and 
the Linder-Huterer model) the fact that matter power spectrum has instabilities
in our model does not imply automatically that such problems will always be present in 
these other UDM framework. Even in the case of GCG the divergences can be cured by 
making sure that the effective speed of sound squared is not negative, as shown 
e.g. in \cite{Amendola2005} and commented on in \cite{Amendola2010}.
Such a construction could also work in an extended model of our type, as
is already indicated in Ref. \cite{Aurich2016} in the case of the arctan-UDM
model.

One merit of the present work is to show that it is possible to learn about 
general features of unified dark sector models from analytical calculations. In our view, 
the physical intuition gained in the process is an important result of our work.




\begin{acknowledgments}

The authors thank Elisa G. M. Ferreira and Guilherme Franzmann for useful discussions. RRC
is grateful to CAPES-Brazil (Grant 88881.119228/2016-01) and LGM acknowledges
CNPq-Brazil (Grant 112861/2015-6) for financial support. EMM is supported
by CAPES-Brazil via IFT/Unesp (Brazil). RB is supported in part by an NSERC Discovery
grant and by the Canada Research Chair program.

\end{acknowledgments}




\appendix*

\section{Background Results \label{app:Background}}  

Inserting Eq.~(\ref{eq:awconstant}) into (\ref{eq:calHwconstant}) leads to:
\begin{equation}
\mathcal{H}\left(\eta\right)=\frac{2}{\left(1+3w\right)}\frac{1}{\left[\frac{2}{\left(1+3w\right)}\frac{1}{\left(H_{0}a_{\text{\text{ref}}}\Omega_{\text{ref}}^{1/2}\right)}+\left(\eta-\eta_{\text{ref}}\right)\right]}\,.\label{eq:calH(eta_ref)wconstant}
\end{equation}
Eqs.~(\ref{eq:awconstant}) and (\ref{eq:calH(eta_ref)wconstant}) for $a(\eta)$ and 
$\mathcal{H}\left(\eta\right)$ depend on three constants: $a_{\text{ref}}$, $\eta_{\text{ref}}$
and $\Omega_{\text{ref}}$. Two of them, say $a_{\text{ref}}$ and
$\eta_{\text{ref}}$, are determined by imposing that the background
evolution is continuous for all values of time. The remaining constant
$\Omega_{\text{ref}}$ is fixed from physically meaningful boundary
conditions. In fact, the continuity equation for
the background, 
\begin{equation}
\tilde{\varepsilon}+3\mathcal{H}\left(\tilde{\varepsilon}+\tilde{p}\right)=0,
\end{equation}
with a barotropic EoS of constant $w$ reads:
\begin{equation}
\Omega\left(a\right)=\Omega_{\text{ref}}\left(\frac{a_{\text{ref}}}{a}\right)^{3\left(1+w\right)}\qquad\left(w=\text{constant}\right)\,.\label{eq:Omegawconstant}
\end{equation}
The matching condition at the CDM-to-DE transition is that the CDM density
parameter $\Omega_{\text{CDM}}\left(\eta_{C}\right)$ calculated at the
transition time $\eta_{C}$ should be the same as DE density parameter
$\Omega_{\text{DE}}\left(\eta_{C}\right)$. This is true because in
the unified matter approach there exists a single fluid that transforms
itself from matter to dark energy, preserving the same energy budget
in both phases. Let
\begin{equation} \Omega_{C}\equiv\Omega_{\text{CDM}}\left(\eta_{C}\right)=\Omega_{\text{DE}}\left(\eta_{C}\right) \, .
\end{equation}
This is the natural choice for energy density reference value $\Omega_{\text{ref}}$:
\begin{equation}
\Omega_{\text{ref}}=\Omega_{C}=\Omega_{\text{DE},0}\left(1+z_{C}\right)^{3\epsilon}\qquad\left(\text{CDM-to-DE transition}\right)\,.\label{eq:OmegaC(OmegaDE)}
\end{equation}
where 
\begin{equation}
\Omega_{\text{DE}}\left(\eta_{0}\right) \equiv \Omega_{\text{DE},0} \, .
\end{equation}
The last equality in Eq.~(\ref{eq:OmegaC(OmegaDE)}) follows from 
Eq. (\ref{eq:Omegawconstant}) applied
to the DE phase $\left(\eta_{C}\leqslant\eta\leqslant\eta_{0}\right)$,
when the EoS parameter is $w=\left(-1+\epsilon\right)$ . Let $a_{\text{CDM}}(\eta)$
[$a_{\text{DE}}(\eta)$] be the scale factor during CDM [DE] regime; and let us use
a similar notation for $\mathcal{H}\left(\eta\right)$. By imposing
\begin{equation}
a_{\text{CDM}}\left(\eta_{C}\right)=a_{\text{DE}}\left(\eta_{C}\right)
\end{equation}
and 
\begin{equation}
\mathcal{H}_{\text{CDM}}\left(\eta_{C}\right)=\mathcal{H}_{\text{DE}}\left(\eta_{C}\right),
\end{equation}
then one gets the remaining constants $a_{\text{ref}}$ and $\eta_{\text{ref}}$.
They become
\begin{equation}
\Omega_{\text{ref}}=\Omega_{C}\,,\qquad\eta_{\text{ref}}=\eta_{C}\,,\qquad a_{\text{ref}}=a_{C}\qquad\left(\text{CDM regime}\right)\,,\label{eq:const_CDM}
\end{equation}
(which is very natural and consistent, by the way) under the assumption
\begin{equation}
\Omega_{\text{ref}}=\Omega_{\text{DE},0}\,,\qquad\eta_{\text{ref}}=\eta_{0}\,,\qquad a_{\text{ref}}=a_{0}=1\qquad\left(\text{DE regime}\right)\,.\label{eq:const_DE}
\end{equation}

Similar considerations can be made at the radiation-to-matter transition. The whole 
picture is the set of equations below which summarize piecewise-UDM background cosmology:
\begin{equation}
a\left(\eta\right)=\begin{cases}
a_{\text{eq}}\left[1-\left(\frac{a_{C}}{a_{\text{eq}}}\right)^{\frac{1}{2}}\left(\frac{a_{0}}{a_{C}}\right)^{\frac{\left(3\epsilon-2\right)}{2}}H_{0}a_{0}\left(\Omega_{\text{DE},0}\right)^{1/2}\left(\eta_{\text{eq}}-\eta\right)\right]\,, & \eta \leqslant \eta_{\text{eq}}\\
a_{C}\left[1-\left(\frac{a_{0}}{a_{C}}\right)^{\frac{\left(3\epsilon-2\right)}{2}}\frac{1}{2}H_{0}a_{0}\left(\Omega_{\text{DE},0}\right)^{1/2}\left(\eta_{C}-\eta\right)\right]^{2}\,, & \eta_{\text{eq}} < \eta \leqslant \eta_{C}\\
a_{0}\left[1-\frac{\left(3\epsilon-2\right)}{2}H_{0}a_{0}\left(\Omega_{\text{DE},0}\right)^{1/2}\left(\eta_{0}-\eta\right)\right]^{\frac{2}{\left(3\epsilon-2\right)}}\,, & \eta_{C} < \eta \leqslant \eta_{0}
\end{cases}\label{eq:a_piecewise}
\end{equation}
with
\begin{equation}
a_{C}=a_{0}\left[1-\frac{\left(3\epsilon-2\right)}{2}H_{0}a_{0}\left(\Omega_{\text{DE},0}\right)^{1/2}\left(\eta_{0}-\eta_{C}\right)\right]^{\frac{2}{\left(3\epsilon-2\right)}}\label{eq:aC}
\end{equation}
and
\begin{equation}
a_{\text{eq}}=a_{C}\left[1-\left(\frac{a_{0}}{a_{C}}\right)^{\frac{\left(3\epsilon-2\right)}{2}}\frac{1}{2}H_{0}a_{0}\left(\Omega_{\text{DE},0}\right)^{1/2}\left(\eta_{C}-\eta_{\text{eq}}\right)\right]^{2}\,,\label{eq:a_eq}
\end{equation}
such that $a\left(\eta\right)$ is indeed continuous throughout the
cosmic history. Also:
\begin{equation}
{\cal H}\left(\eta\right)=\frac{2}{\left(1+3w\right)}\frac{1}{\bar{\eta}}\,,\label{eq:calH_piecewise}
\end{equation}
where
\begin{equation}
\bar{\eta}\left(\eta\right)=\begin{cases}
\bar{\eta}_{\gamma}\left(\eta\right) \equiv \frac{1}{2}\bar{\eta}_{\text{CDM}}\left(\eta_{\text{eq}}\right) - \left( \eta_{\text{eq}} - \eta \right) \,, & \eta \leqslant \eta_{\text{eq}}\\
\bar{\eta}_{\text{CDM}}\left(\eta\right) \equiv \left(3\epsilon-2\right)\bar{\eta}_{\text{DE}}\left(\eta_{C}\right) - \left( \eta_{C} - \eta \right) \,, & \eta_{\text{eq}} < \eta \leqslant\eta_{C}\\
\bar{\eta}_{\text{DE}}\left(\eta\right)\equiv\frac{2}{\left(3\epsilon-2\right)}\frac{1}{H_{0}a_{0}\left(\Omega_{\text{DE},0}\right)^{1/2}}-\left(\eta_{0}-\eta\right)\,, & \eta_{C} < \eta \leqslant \eta_{0}
\end{cases}\label{eq:etabar_piecewise}
\end{equation} 
with
\begin{equation}
\bar{\eta}_{\gamma}\left(\eta\right)=\bar{\eta}_{\text{CDM}}\left(\eta\right)-\frac{1}{2}\bar{\eta}_{\text{CDM}}\left(\eta_{\text{eq}}\right)\label{eq:etabar_gamma}
\end{equation}
and
\begin{equation}
\bar{\eta}_{\text{CDM}}\left(\eta\right)=\bar{\eta}_{\text{DE}}\left(\eta\right)+3\left(-1+\epsilon\right)\bar{\eta}_{\text{DE}}\left(\eta_{C}\right)\,.\label{eq:etabar_CDM}
\end{equation}
Note how ${\cal H}\left(\eta\right)$ is also continuous at $\eta_{\text{eq}}$
and $\eta_{C}$. 

The first line in Eq. (\ref{eq:a_piecewise}) is precisely the expression
needed for writing the last equality in Eq.~(\ref{eq:deltaminusSS}). 

Eqs.~(\ref{eq:etabar_gamma}) and (\ref{eq:etabar_CDM}) show that $\bar{\eta}$
is discontinuous both at $\eta=\eta_{\text{eq}}$ and $\eta=\eta_{C}$.
This happens in order to compensate for the discontinuity introduced by
our piecewise $w$ \textemdash{} see Eq.~(\ref{eq:w_piecewiseUDM}) \textendash{} and makes
it possible that $a\left(\eta\right)$ and ${\cal H}\left(\eta\right)$
are continuous. The discontinuity in $\bar{\eta}$ is not a problem because
this quantity is defined only for convenience; it does not have
an intrinsic physical meaning. The  convenience is that the perturbation
equation for $\phi_{{\bf k}}$ assumes the form (\ref{eq:DiffEqphi(eta)wconstant})
in terms of $\bar{\eta}$; and this form is free from subtleties regarding integration
constants from the physics of the background cosmology
\footnote{Indeed, Eq.~(\ref{eq:calH(eta)wconstant})
should be understood as Eq.~(\ref{eq:calH_piecewise}).} .
Although these subtleties are not important for the technical task
of obtaining the solution of (\ref{eq:DiffEqphi(eta)wconstant}),
they are relevant when one performs the matching of different phases according to 
matching conditions.
This explains the reason for the factor $(3\epsilon-2)\eta_C$
in the expression for $\phi_{\mathbf{k}-}(\eta_C)$, Eq.~(\ref{eq:phi_minus(etaC)}).




\end{document}